﻿
\documentclass[aip,pop,reprint]{revtex4-1}

\usepackage{lineno}
\usepackage{graphicx}

\begin{document}

\title{Validation and results of an approximate model for the stress of a Tokamak toroidal field coil at the inboard midplane}

\author{C.P.S. Swanson}
\email{cswanson@pppl.gov}
\affiliation{Princeton Plasma Physics Laboratory, Princeton University, Princeton, New Jersey 08543, USA}
\author{S. Kahn}
\affiliation{Culham Centre for Fusion Energy, Culham, Oxfordshire, UK}
\author{C. Rana}
\affiliation{Princeton Plasma Physics Laboratory, Princeton University, Princeton, New Jersey 08543, USA}
\author{P.H. Titus}
\affiliation{Princeton Plasma Physics Laboratory, Princeton University, Princeton, New Jersey 08543, USA}
\author{A.W. Brooks}
\affiliation{Princeton Plasma Physics Laboratory, Princeton University, Princeton, New Jersey 08543, USA}
\author{W. Guttenfelder}
\affiliation{Princeton Plasma Physics Laboratory, Princeton University, Princeton, New Jersey 08543, USA}
\author{Y. Zhai}
\affiliation{Princeton Plasma Physics Laboratory, Princeton University, Princeton, New Jersey 08543, USA}
\author{T.G. Brown}
\affiliation{Princeton Plasma Physics Laboratory, Princeton University, Princeton, New Jersey 08543, USA}
\author{J.E. Menard}
\affiliation{Princeton Plasma Physics Laboratory, Princeton University, Princeton, New Jersey 08543, USA}

\begin{abstract}

We present the verification, validation, and results of an approximate, analytic model for the radial profile of the stress, strain, and displacement within the toroidal field (TF) coil of a Tokamak at the inner midplane, where stress management is of the most concern. The model is designed to have high execution speed yet capture the essential physics, suitable for scoping studies, rapid evaluation of designs, and in the inner loop of an optimizer. It is implemented in the PROCESS fusion reactor systems code. The model solves a many-layer axisymmetric extended plane strain problem. It includes linear elastic deformation, Poisson effects, transverse-isotropic materials properties, radial Lorentz force profiles, and axial tension applied to layer subsets. The model does not include out-of-plane forces from poloidal field coils. We benchmark the model against 2D and 3D Finite Element Analyses (FEA) using Ansys and COMSOL. We find the Tresca stress accuracy of the model to be within 10\% of the FEA result. We show that this model allows PROCESS to optimize a fusion pilot plant, subject to the TF coil winding pack and coil case yield constraints. This model sets an upper limit on the magnetic field strength at the coil surface of $29$ Tesla for steel TF coil cases, with the practical limit being significantly below this. 

\end{abstract}

\maketitle

\section{Introduction}

Finite Element Analysis (FEA) tools such as Ansys\cite{noauthor_ansys_nodate} and COMSOL\cite{noauthor_comsol_nodate} enjoy great success in industry, and are commonly used to model the designs of future Tokamak-based fusion power plants. However, these large structural FEA models can take hours to execute and produce results. Before a design rises to the level of resolution and fidelity that warrant a finite element analysis, it may be analyzed and adjusted many times. Approximate, fast models are needed for this medium-fidelity phase in the design. This paper describes the verification, validation, and results of an approximate model for the stress and strain distribution at the inner midplane of a TF coil, suitable for rapid iteration of a medium-fidelity design or in the inner loop of a numerical optimizer.

Compared to the ITER experiment under construction, Tokamak-based fusion power plants will require operation with increased structural load (forces and stresses) on the inner leg of the toroidal field (TF) coil. One reason for this is higher magnetic field amplitudes of the approach espoused by Commonwealth Fusion Systems (CFS) in their proposed SPARC experiment and ARC reactor\cite{greenwald_high-field_2018,greenwald_performance_2018,sorbom_arc_2015}, and earlier the earlier proposed experiments IGNITOR and Compact Ignition Tokamak (CIT).\cite{coppi_optimal_2001,flanagan_overview_1986} Another design philosophy leading to increased stress in the TF coil proposes a reactor much larger than ITER, espoused by the EU-DEMO and K-DEMO.\cite{federici_demo_2018,corato_progress_2018,kim_design_2015} Finally, another design philosophy is that of the Spherical Tokamak, whose TF coil system has a very small cross sectional area at the inner midplane. A Spherical Tokamak reactor design is described by Tokamak Energy Ltd.\cite{costley_towards_2019} and supported by such experiments as NSTX-U and MAST.\cite{menard_overview_2012,morris_mast_2012} 

The benefit of a fast, approximate method of computing the stress and strain distribution is that it can be used to rapidly iterate through many facility designs. Codes that do this are referred to as systems codes. They couple models from many different fields, from plasma physics to engineering and economics. For example, the PROCESS systems code can adjust the magnetic field strength to affect the fusion power (plasma physics), to optimize the levelized cost of electricity (economic), subject to materials yield constraints (engineering). Numerically, PROCESS runs by adjusting some number of ``iteration variables" to optimize a figure of merit and satisfy some number of constraints. PROCESS analyses typically use a few dozen iteration variables and constraints.

There are many systems codes in the literature, such as Culham Centre For Fusion Energy (CCFE) PROCESS,\cite{kovari_process_2014,kovari_process_2016} Tokamak Energy's TESC,\cite{costley_power_2015,costley_towards_2019} General Atomics's GASC,\cite{stambaugh_fusion_2011,buttery_advanced_2021}, ORNL's Unnamed FESS Systems code,\cite{kessel_core_2018,kessel_fess_2021} and CEA's SYCOMORE.\cite{reux_demo_2015} For a very tutorial approach, see papers from J. Freidberg's group.\cite{freidberg_designing_2015,segal_steady_2021,djsegal_djsegalfusionsystemsjl_2020} 

The present work represents an increase in the accuracy of the model for TF coil stress over these systems codes. The Freidberg systems studies evaluate only the axial (tension) stress in the TF coil. GASC evaluates the axial tension stress and the toroidal compression (wedge) stress according to a simple lumped-element model. A two-layer spreadsheet model by Titus \textit{et al.} includes the effect of differences in Young's modulus between the winding pack and structure.\cite{titus_fnsf_2021} These models miss effects which are captured by the present model and can be $O(1)$, for example Poisson effects, anisotropy of the winding pack elastic properties, the radially-resolved volumetric Lorentz force density, and structures which contain $>2$ layers.

The structure of the remainder of this paper is as follows: Section \ref{sec:Overview} gives an overview of the present model, describing its most important features. Section \ref{sec:Implications} discusses some of the general implications of the model on TF coil design. 

Sections \ref{sec:Simple} and \ref{sec:Complex} represent ``Verification and Validation" of the model, in the sense that ``verification" verifies that the numerical methods are sufficient to solve the approximate model, and ``validation" validates that the approximate model is a sufficiently accurate model of the full system. Section \ref{sec:Simple} presents the verification, the results of FEA analyses. Section \ref{sec:Complex} presents validation, the results of complex, 3D benchmarks, evaluating the accuracy of the model when applied to full facility designs from the literature. 

Section \ref{sec:Optimize} demonstrates the use the model in the optimization of a facility design: a facility design from the literature is optimized for net electric power. PROCESS used the model to enforce the Tresca yield constraint at the TF coil case and winding pack. Section \ref{sec:Conclusion} discusses the findings and concludes.

\section{Overview of the approximate model}
\label{sec:Overview}

In this paper, we present the verification, validation, and results of an approximate model for the stress and strain of the TF coil at the inboard midplane. This model has recently been implemented in the PROCESS fusion reactor systems code. In this paper, we only describe a broad overview of the model; for a detailed description and discussion see our companion paper\cite{kahn_derivation_2022} (in preparation) or the writeup in the supplemental material.\cite{swanson_supplemental_2022}

For a description of the previous TF coil stress model, which this model replaces, see Reference \cite{morris_implications_2015}.

\begin{figure}
\begin{center}
\includegraphics[width=0.6\linewidth]{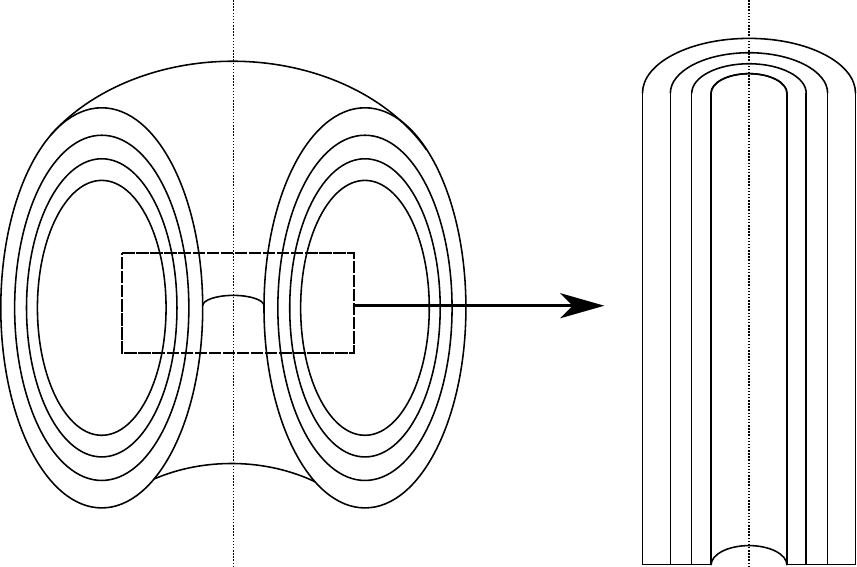}
\caption{The assumed geometry of the approximate model: The inboard midplane of the TF coil system is assumed to map to a multilayer axisymmetric cylinder, long in $\hat{z}$, the axial direction.}
\label{fig:Tok2Cyl}
\end{center}
\end{figure}

The model is an extension of the Lam\'e thick cylinder pressure vessel solutions.\cite{harvey_theory_1985,iit_kharagpur_design_2009} The stress, strain, and displacements are assumed to follow linear elastic deformation. The TF coil structure is assumed to be axisymmetric and long in the axial direction; see Figure \ref{fig:Tok2Cyl}.

The model deviates from the Lam\'e thick cylinder pressure vessel model in a number of important ways:

The TF coil structure is assumed to be under axial tension. This is a result of the vertical separating force of the TF coil system caused by the Lorentz force on the top and bottom of the TF coils. The vertical separating force is calculated as part of a separate PROCESS model and is considered an input to this model. This model assumes that the axial strain $\epsilon_z$ is uniform and finds the self-consistent value of $\epsilon_z$ that produces the prescribed vertical separating force. In the diction of Cheng \textit{et al.}, this is an ``extended plane strain problem".\cite{cheng_generalized_1995}

The model is formulated with respect to an arbitrary number of layers, rather than one thick cylindrical shell. Elastic properties within each layer are assumed to be uniform, but are different for each layer. In this manner, many components within the TF coil system may be resolved: The structure of the Central Solenoid (ex: steel), a soft inter-layer (ex: Kapton), the inner TF coil case (ex: steel), the winding pack (composite of steel, void, copper, and superconductor), and the outer TF coil case (ex: steel). Depending on the thickness of the inter-layer and the stiffness of the winding pack, the stress profile deviates widely from the Lam\'e thick cylinder pressure vessel profile. 

The model resolves the radial profile of the radial $f(r) = j_z(r) B_\theta(r)$ force density, rather than collapsing it into a scalar pressure. Depending on the thickness and the stiffness of the winding pack, this can be a large effect.

Poisson effects are fully resolved. This causes interaction between the compressive wedge and radial pressures and the axial tension, concentrating the axial tension at the outer radii. This can be a leading order effect.

The Young's modulus $E$ and Poisson's ratio $\nu$ may be transverse-isotropic rather than isotropic. This is most important for the winding pack, where a honeycomb of steel conduit or jacket provides very different elastic properties in the axial and transverse directions. 

The Young's modulus and Poisson's ratios for an effective, homogenized winding pack are computed from an assumed 2D winding pack cross sectional layout. Likewise, the stress in each individual component of the winding pack are computed after the homogenized solution is found. These computations are known as smearing and unsmearing, respectively. The details of the smearing and unsmearing algorithms can be found in the supplemental material.\cite{swanson_supplemental_2022}

The smearing and unsmearing algorithms are extensions of the Voigt rule and the Reuss rule\cite{voigt_ueber_1889,reuss_berechnung_1929,noauthor_derivation_nodate} formulated for transverse-isotropic materials. They are formulated in the compliance form, with the assumption that the boundary conditions of the transverse structure are free to deform unimpeded by the deformation of neighboring structure. This is necessarily an approximation. The compliance form can be readily inverted into the stiffness form, as the Poisson's ratios are also computed.

For the axial Young's modulus $E_z$ and the axial-transverse Poisson's ratio $\nu_{z,\perp}$, only the relative fractions of insulation, conduit, copper, superconductor, and void are required. Parallel-composition via the generalized Voigt rule is used.

\begin{figure}
\begin{center}
\includegraphics[width=0.95\linewidth]{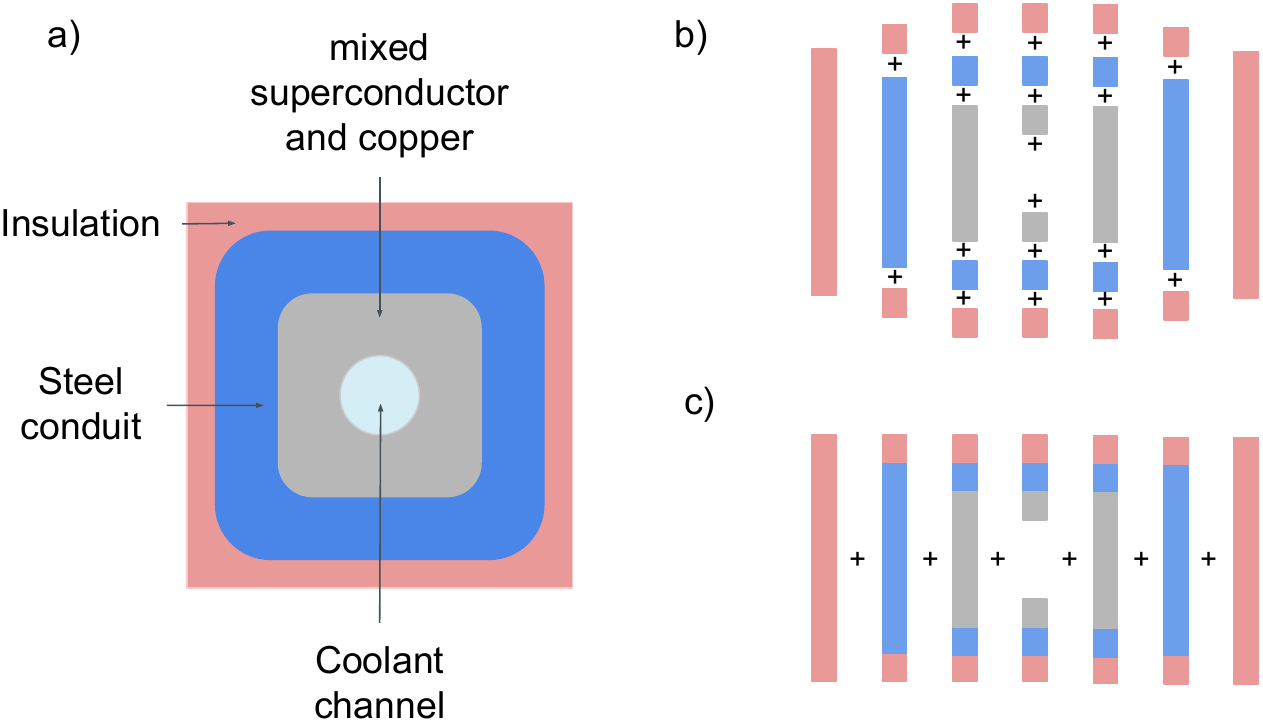}
\caption{Illustration of the winding pack smearing algorithm. a) The assumed 2D cross sectional layout of a single turn of the winding pack: Nested squares of insulation, conduit, conductor and copper, and coolant channel. b) The series-composition step: Each component of each lamina of the winding pack has its Young's modulus and Poisson's ratio series-composited. c) The parallel-composition step: All laminae are parallel-composited together.}
\label{fig:wp_decomposition}
\end{center}
\end{figure}

For the transverse Young's modulus $E_\perp$ and the transverse-transverse Poisson's ratio $\nu_\perp$, a 2D winding pack structure must be assumed. The winding pack is assumed to be a regular grid of the unit cell shown in Figure \ref{fig:wp_decomposition} (a). As shown in Figure \ref{fig:wp_decomposition} (b), the unit cell is decomposed into parallel laminae, and each lamina is decomposed into rectangular-cross-section components of pure materials. The components are series-composited together according to the generalized Reuss rule. As shown in Figure \ref{fig:wp_decomposition} (c), the laminae are then parallel-composited together according to the generalized Voigt rule. This is necessarily an approximation, as this model admits relative deformation of laminae, where in a real winding pack the shear modulus impedes this. 

While some details of the cabling strategy are not resolved by this model, they may be approached. If the conductor is composed of free filaments with a large void fraction, the transverse Young's modulus of the conductor may be set to zero. If the conductor is axially stiff but is wound with significant twist pitch, the axial Young's modulus of the conductor may be set to zero or de-rated by some factor. In this way, cabling strategies such as stacked tape and cable-in-conduit conductor (CICC) may be approximately differentiated. 

The model also considers the case that some inner subset of layers are under a different (or zero) axial tension, such as when the central solenoid and/or bucking cylinder is decoupled from the TF coil via a frictionless interlayer. This is the case when the TF coil system is ``bucked and wedged."\cite{titus_structural_2003} 

PROCESS uses the Tresca criterion to model material yield thresholds. In words, the Tresca criterion models every failure as a shear failure. Materials are deemed to have failed when the following quantity exceeds a certain threshold:
\begin{equation}
\sigma_{Tresca} = \textrm{max}(|\sigma_r-\sigma_\theta|,|\sigma_r-\sigma_z|,|\sigma_\theta-\sigma_z|)
\label{eq:Tresca}
\end{equation}
where $\sigma_r,\sigma_\theta,\sigma_z$ are the stresses in the radial, azimuthal, and axial directions, and $\sigma_{Tresca}$, the ``Tresca stress," indicates material yield when it exceeds a specific value.

Note that the radial ``centering" force of a TF coil causes $\sigma_r,\sigma_\theta<0$ and the vertical separating force of a TF coil causes $\sigma_z>0$. Typical thresholds for $\sigma_{Tresca}$ are in the range 600 - 670 MPa for structural steel. 

Several features are missing from this model. The out-of-plane forces caused by the interaction with the TF and Poloidal Field (PF) coils is not resolved. Real TF coil systems are made of 10 - 20 individual TF coils, not a continuous, axisymmetric coil. The individual TF coils have non-axisymmetric components, including lateral sidewalls around the winding pack. For these reasons, a 10\% agreement with a full FEA analysis would be considered satisfactory.

\section{Broad implications of the model}
\label{sec:Implications}

This section discusses the broad implications of the model, which may be used to guide intuition and produce design heuristics.

\subsection{Uniform transverse stress in a uniform, solid cylinder under external pressure}
\label{sec:Hydrostatic}

This case considers a solid cylinder with uniform materials properties, not a shell. This corresponds to the case that the inner leg of the TF coil is bucked against a solid (not hollow) bucking cylinder, or the case that the central column of a Spherical Tokamak is solid and has no central solenoid. This is sometimes called a ``plugged" bore. We consider some finite axial force $F_z$ and some finite external pressure $P_{ext}$. According to the Lam\'e thick cylinder pressure vessel solutions, the solution to the transverse stress distribution is $\sigma_r = \sigma_\theta = -P_{ext}$ everywhere within the cylinder. This is analogous to the hydrostatic-like case, where a uniform external pressure produces a uniform internal pressure. 

Note that while $\sigma_r = \sigma_\theta < 0$, the axial stress $\sigma_z$ is in general different and positive, so failure still occurs for high enough $P_{ext}$ under the Tresca criterion.

\subsection{Thin current-carrying layer acts like an external pressure}

\begin{figure}
\begin{center}
\includegraphics[width=0.99\linewidth]{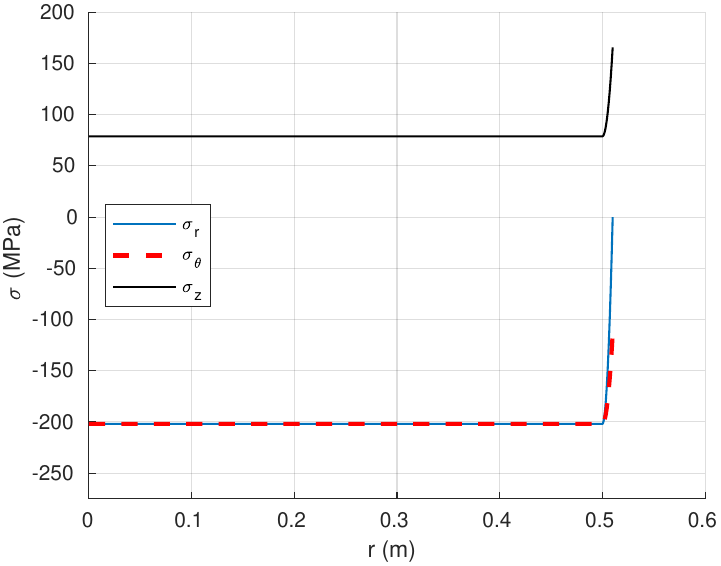}
\caption{The radial stress profile $\sigma_{r,\theta,z}(r)$ resulting from the case of a uniform cylinder with a thin current-carrying layer on the outside. The current is sufficient to make $B=22.4$ Tesla, $P_{mag}=200$ MPa. As expected, other than the thin current-carrying layer (1 cm), $\sigma_r = \sigma_\theta = -P_{ext}$ everywhere within the cylinder. $\sigma_z$ is due to a finite $F_z$. $\sigma_z$ is concentrated in the outer layer due to the finite Poisson ratio, $\nu=0.3$.}
\label{fig:PluggedThinLorentz}
\end{center}
\end{figure}

This case replaces the external pressure, above, with a thin current-carrying layer, carrying sufficient current to cause the external magnetic pressure $P_{mag}$ to be equal to the assumed external pressure, $P_{mag}=B^2/2\mu_0 = P_{ext}$. We would expect the external pressure case to be recovered as the thickness of the current-carrying layer $\rightarrow 0$, and indeed it does according to Figure \ref{fig:PluggedThinLorentz}.

Incidentally, the distribution of $\sigma_z$ is shown to concentrate in the current-carrying layer. This is due to the finite Poisson's ratio assumed $\nu=0.3$. Where $\sigma_{r,\theta}<0$, the material is being transversely squeezed and expands in the axial direction, relieving $\sigma_z$ there. $\sigma_z>0$ is caused by an assumed axial tension, $F_z$.

Note that a winding pack which is thick and stiff must be evaluated using the full self-consistent model, as it partially supports itself with wedge pressure.

\subsection{A finite bore causes a large $\sigma_\theta$}
\label{sec:FiniteBore}

This case considers a uniform cylindrical shell with finite bore ($r_1>0$), under an external pressure $P_{ext}$. The thick cylinder Lam\'e pressure vessel solution predicts the value of the azimuthal stress on the inner edge, $\sigma_\theta(r_1)$. It is:
\begin{equation}
\sigma_\theta(r_1) = -2P_{ext} \frac{1}{1-r_1^2/r_2^2}
\label{eq:LameInnerHoopStress}
\end{equation}
and indeed the model reproduces this.

\subsection{Azimuthal stress $\sigma_\theta$ is usually limiting}

Recall that Equation \ref{eq:LameInnerHoopStress} holds when there is a uniform cylindrical shell under external pressure, or equivalently with a thin current-carrying layer on the outside. 

Equation \ref{eq:LameInnerHoopStress} is worth studying, because $\sigma_\theta(r_1)$ is often the largest compressive stress in the cross section, and therefore, combined with the axial tension stress $\sigma_z$ determines whether the coil will fail. In the thin-shell limit, $r_2 = r_1 + \delta r$, $\sigma_\theta(r_1) \rightarrow -P_{ext} \frac{r_1}{\delta r}$, much more stress than $P_{ext}$. In the thick-shell limit, $r_2\gg r_1$, $\sigma_\theta(r_1) \rightarrow -2P_{ext}$, less than the thin-shell case but still double the stress of the solid-cylinder case. 

It is notable that the limiting behavior of $r_1\rightarrow 0$ is not the same as $r_1=0$. As we saw in the solid-cylinder case, $\sigma_\theta(0) = -P_{ext}$ when $r_1=0$, only half the stress of the small-but-finite bore case. Even a small hole causes stress concentration in a bulk material.

\subsection{Stress is concentrated in the stiffest member}

This case considers a 3-layer TF coil: An inner layer with a low Young's modulus, a middle layer with a high Young's modulus, and an outer layer with a low Young's modulus. The coil is under a finite external pressure $P_{ext}$, or equivalently contains a very thin current-carrying layer at the outer edge. 

\begin{figure}
\begin{center}
\includegraphics[width=0.99\linewidth]{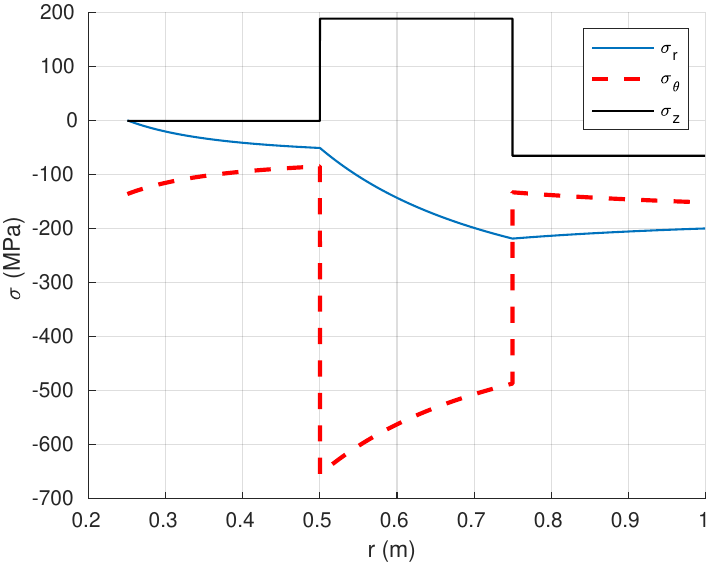}
\caption{The radial stress profile $\sigma_{r,\theta,z}(r)$ resulting from the case of a three-layer cylinder with the middle layer $10\times$ stiffer than the inner and outer layer. As expected the axial and azimuthal stress $\sigma_{z,\theta}$ are concentrated in the stiffest layer.}
\label{fig:ShellStiffConcentration}
\end{center}
\end{figure}

We would expect the axial and azimuthal stress $\sigma_z,\sigma_\theta$ to be concentrated in the middle, stiffest layer, and indeed that is what the model produces. See Figure \ref{fig:ShellStiffConcentration}. The $\sigma_r$ case is more complex, as its boundary conditions must be $0$ at the inner edge and $-P_{ext}$ at the outer edge. 

This is a notable finding: A broad, stiff layer may alleviate stress from delicate components such as the superconducting winding pack. On the other hand, care must be taken to prevent this stiffer layer from exceeding its yield criterion.

\subsection{Reducing $F_z$}
\label{sec:Tensionless}

Of the three differences in the Tresca criterion (Eq. \ref{eq:Tresca}), usually the limiting one is $|\sigma_\theta-\sigma_z|$. Since typically $\sigma_\theta<-P_{mag}$, one school of TF coil design holds that it is essential to reduce $\sigma_z$, and therefore $F_z$, the axial tension. This is the portion of the vertical separating force taken up by the inner leg of the TF coil. This has been achieved by several methods in the literature and in practice, including combinations of compression rings at the top and bottom of the inner leg, slip/sliding joints at the same location which do not transmit force, and even hydraulic pre-compression structures which compress the inner leg.\cite{titus_structural_2003,flanagan_overview_1986}

\subsection{Magnetic fields corresponding to steel yield criteria}

A common threshold to determine structural steel failure based on Equation \ref{eq:Tresca} is 660 MPa. This sets limits on the strength of magnetic field at the outer surface of the TF coil inner leg which can be achieved. 

The most generous scenario is the case that $F_z$ may be reduced significantly by a combination of methods discussed in Section \ref{sec:Tensionless}, and that there is no bore (no internal central solenoid for plasma startup or heating). This corresponds to a solid steel bucking cylinder (a ``plugged bore", Section \ref{sec:Hydrostatic}) rather than a coil case with a finite bore. This scenario may be evaluated using the condition discussed in Section \ref{sec:Hydrostatic}, $\sigma_\theta = -P_{mag}$. The highest magnetic field achievable using a solid steel inner cylinder is therefore that which corresponds to $P_{mag}=660$ MPa, $B = \sqrt{2\mu_0 P_{mag}} \approx 40.7$ Tesla. At a higher magnetic field than this, the inner bucking cylinder will yield.

However, as soon as a bore is added, the allowable $P_{mag}$ for a steel case halves to 330 MPa in accordance with Equation \ref{eq:LameInnerHoopStress} in Section \ref{sec:FiniteBore}, and decreases further as the radius of the bore increases to accommodate startup flux (Volt-seconds). In this case, the magnetic field limit for a steel coil case is $B < 28.8$ Tesla. The 23 - 25 Tesla design of SPARC and ARC approaches this limit. 

\section{FEA verification of the model in simple geometries}
\label{sec:Simple}

This section ``verifies" the model under the definition that ``verification" establishes that the numerical methods are sufficient to accurately solve the approximate model. Establishing that the approximate model is sufficiently accurate to the full system is here called ``validation" and is discussed in Section \ref{sec:Complex}.

\begin{figure}
\begin{center}
\includegraphics[width=0.7\linewidth]{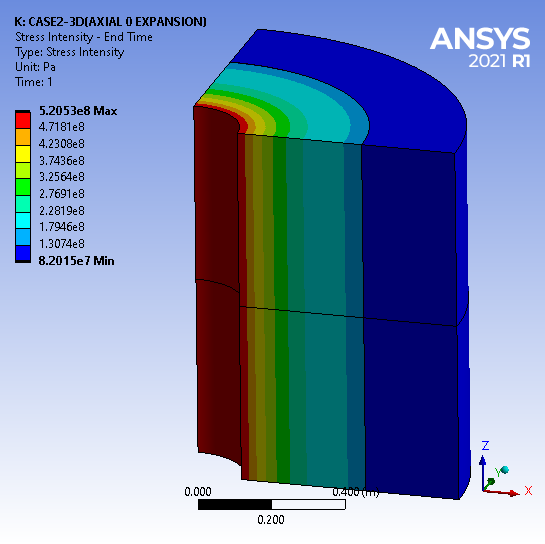}
\caption{An example Ansys computational domain, of which an FEA was run. FEA like this were used to verify the model. The color scale shows the ``stress intensity", equal to the parameter $\sigma_{Tresca}$ defined in Equation \ref{eq:Tresca}}
\label{fig:AnsysSimple}
\end{center}
\end{figure}

The results of the model are compared against 2D and 3D Ansys FEA written by C. Rana. An example of an Ansys computational domain is given in Figure \ref{fig:AnsysSimple}. 

\subsection{Verification}

The first verification tests the present model's treatment of the radial Lorentz body force density. The geometry is a uniform cylindrical shell with inner radius $r_1 = 0.25$ m, outer radius $r_2 = 0.5$ m. The material has an isotropic Young's modulus $E=200$ GPa and an isotropic Poisson's ratio $\nu = 0.3$ ($\sim$ generic structural steel). The axial current density producing the Lorentz force density is assumed to be uniform, $j_z = 95.1$ MA/m$^2$, sufficient to produce $B=22.4$ Tesla, $P_{mag} = 200$ MPa. The axial strain is fixed to be $\epsilon_z=0$, though due to Poisson effects this results in a finite axial compression $F_z<0$. Cases corresponding to axial tension, with $\epsilon_z>0, F_z>0$, were also run but are not shown here.

\begin{figure}
\begin{center}
\includegraphics[width=0.99\linewidth]{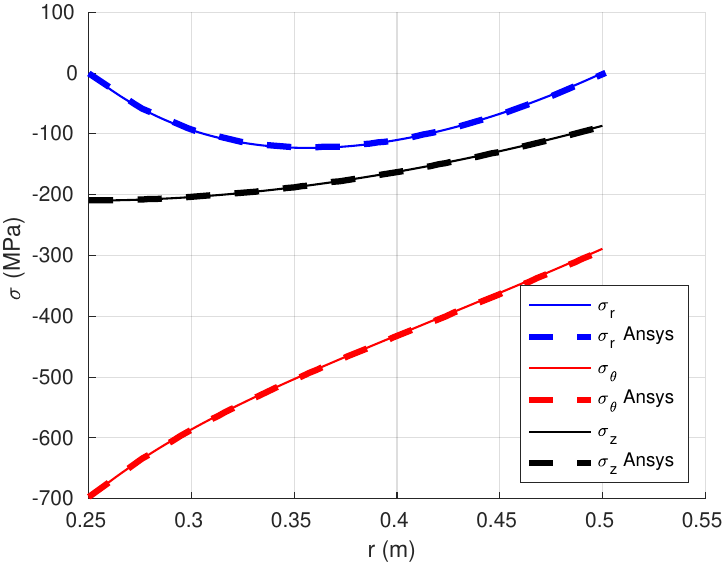}
\caption{The radial stress profile $\sigma_{r,\theta,z}(r)$, according to the model and the Ansys FEA. The simulated case was a uniform, isotropic cylindrical shell with constant axial current density. The Ansys result differs from the model result by at most $0.7\%$, presumably due to the finite resolution of the FEA computational domain.}
\label{fig:S03_Lorentz}
\end{center}
\end{figure}

The equivalent geometry was set up in Ansys, and FEA were run. The FEA was a static 3D structural analysis, with the body force density externally evaluated and imposed on each volume element. One quarter of the azimuthal extent was simulated, with fourfold radial symmetry and axially fixed boundary conditions. The shell was 1 m long, and the upper and lower faces were fixed corresponding to $\epsilon_z=0$. The radial force density was computed according to Ampere's Law and Lorentz's Law, $f(r) \approx 5.69 \textrm{ GN/m}^3 \times (r/1\textrm{m}) - 0.356 \textrm{ GN/m}^3 * (1\textrm{m}/r)$. 

The results are shown in Figure \ref{fig:S03_Lorentz}. The model results agree with the Ansys results to within $0.7\%$, presumably due to the finite resolution of the Ansys FEA computational domain. We have therefore verified that the model treats the Lorentz body force density correctly. 

The second verification tests the model's treatment of anisotropy (transverse-isotropy) and multiple layers. The geometry is two nested cylindrical shells. The inner layer has radius $r_1 = 0.33$ m, outer radius $r_2 = 0.67$ m. The outer layer has inner radius $r_2 = 0.67$ m, outer radius $r_3 = 1.0$m. The inner layer material has an isotropic Young's modulus $E=200$ GPa and an isotropic Poisson's ratio $\nu = 0.3$. The outer layer material is transverse-isotropic, and is less stiff in the \textit{transverse} direction only. It has axial isotropic Young's modulus $E_z=200$ GPa, transverse Young's modulus $E_\perp=100$ GPa, axial-transverse Poisson's ratio $\nu_{z,\perp} = 0.3$, and transverse-transverse Poisson's ratio $\nu_\perp = 0.5$. The exterior is subjected to a pressure $P_{ext} = 200$ MPa. The axial strain is fixed to be $\epsilon_z=0$, though due to Poisson effects this results in a finite axial compression $F_z<0$. $\epsilon_z>0, F_z>0$ cases were also run, but are not shown here.

\begin{figure}
\begin{center}
\includegraphics[width=0.99\linewidth]{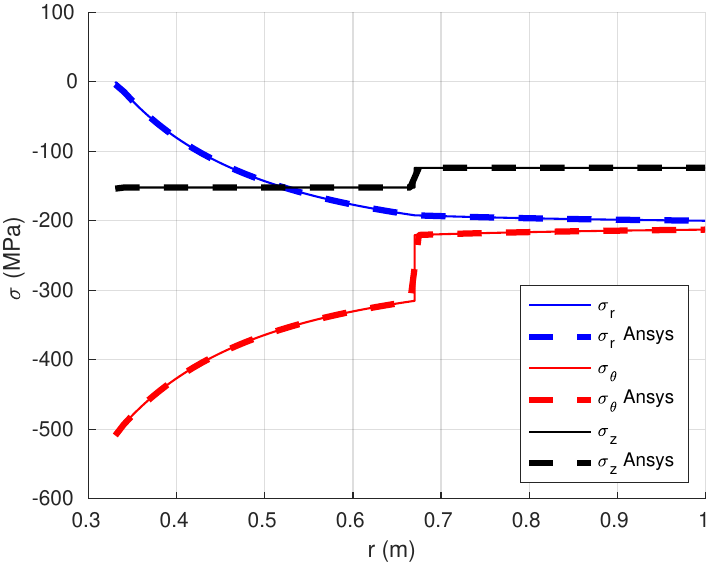}
\caption{The radial stress profile $\sigma_{r,\theta,z}(r)$, according to the model and the Ansys FEA. The simulated case was two nested cylindrical shells, the outer of which has a transverse-isotropic Young's modulus which is less stiff in the transverse direction. The Ansys result differs from the model result by at most $0.3\%$, presumably due to the finite resolution of the FEA computational domain.}
\label{fig:S05_Anisotropic}
\end{center}
\end{figure}

The equivalent geometry was set up in Ansys, and FEA were run. The FEA was a static 3D structural analysis, with an externally imposed pressure. One quarter of the azimuthal extent was simulated, with fourfold radial symmetry and axially fixed boundary conditions. The shell was 1 m long, and the upper and lower faces were fixed corresponding to $\epsilon_z=0$. An external pressure of $P_{ext}=200$ MPa was enforced. 

The results are shown in Figure \ref{fig:S05_Anisotropic}. The model results agree with the Ansys results to within $0.3\%$, again presumably due to the finite resolution of the Ansys FEA computational domain. We have therefore verified that the model treats anisotropy and multi-layered structures correctly. 

\subsection{Quantifying the effect non-axisymmetry}
\label{sec:polygon}

\begin{figure}
\begin{center}
\includegraphics[width=0.6\linewidth]{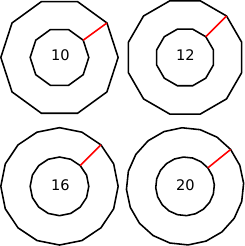}
\caption{The computational domain for the Ansys FEA in Section \ref{sec:polygon}. They are polygonal shells with 10, 12, 16, and 20 sides. This represents a straight-inner-side TF coil system with discrete TF coils. The red line is a meeting between two sectors, where the stress concentration is expected to be the highest. Real TF coil systems are made of discrete TF coil sectors, but the inner case surface is curved to form a sector of a cylinder, rather than straight.}
\label{fig:Polygons}
\end{center}
\end{figure}

The analysis discussed in this section relaxes the assumption that the TF coil is perfectly axisymmetric. The geometry assumed by the model is exactly the same as the two-layer transverse-isotropic validation seen in Figure \ref{fig:S05_Anisotropic}. However, the Ansys FEA now considers a polygonal, rather than cylindrical, TF coil inner leg. The geometry, materials properties, and external pressure are otherwise the same. Polygons with side numbers of 10 - 20 are considered, the typical number of Tokamak TF coils. The geometry is depicted schematically in Figure \ref{fig:Polygons}. 

\begin{figure}
\begin{center}
\includegraphics[width=0.99\linewidth]{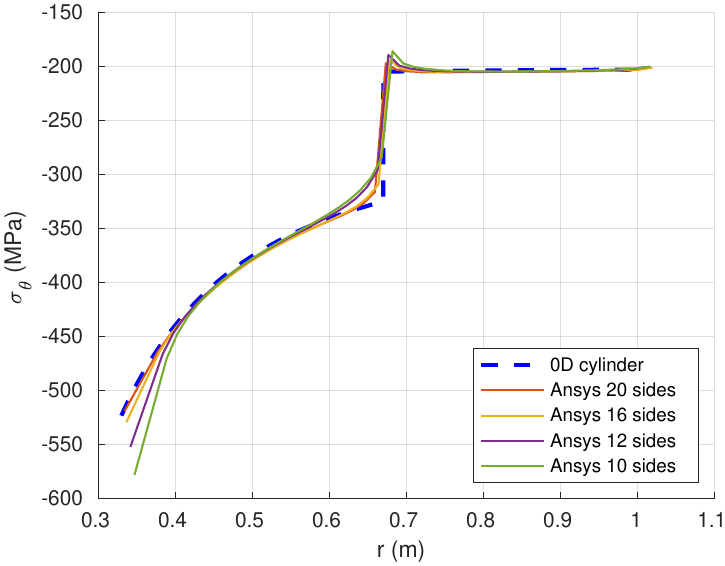}
\caption{The radial profile along the red lines in Figure \ref{fig:Polygons} of the azimuthal stress. The cylinder case of Figure \ref{fig:S05_Anisotropic} is shown, along with Ansys FEA results from 10, 12, 16, and 20-sided polygonal shells. As the side number increases, the stress profile more closely approximates that of a cylinder.}
\label{fig:PolygonAnalysis}
\end{center}
\end{figure}

We expect stress to concentrated at the inside corners of the trapezoidal cross section, where one sector meets the next. One of these meeting lines between sectors is shown in red in each polygon shown in Figure \ref{fig:Polygons}. Therefore, we plot the azimuthal stress distribution ($\sigma_\theta$) along this worst-case stress concentration line. The results are shown in Figure \ref{fig:PolygonAnalysis}. That figure shows that, as the polygon side number increases, the azimuthal stress profile more closely approximates that of a cylinder. 

For 20 TF coils, the the maximum compressive azimuthal stress discrepancy is less than $1\%$. For 16 TF coils, the discrepancy is $1.2\%$. For 12 TF coils, the discrepancy is $5.5\%$. For 10 TF coils, the discrepancy is $10.5\%$. 

We note that a polygonal shell is a worst-case scenario, as real TF coil inner case surfaces are curved to form a sector of a cylinder. This minimizes the stress concentration. 

\section{Validation of the model against full 3D FEA}
\label{sec:Complex}

This section ``validates" the model under the definition that ``validation" establishes that the approximate model is sufficiently accurate to capture the relevant phenomena. The model is validated by comparing it to fully 3D, non-ideal FEA that were performed in the literature using Ansys and COMSOL. The model reproduces the relevant stresses to within 10\%. We do not precisely account for this 10\% discrepancy, but it is deemed sufficiently accurate for broad system studies, suitable for batch runs of systems codes and within an optimizer loop. Candidates for this 10\% discrepancy are non-axisymmetric features such as the lateral side-wall casing, the inherently approximate nature of property-smearing, and 3D features such as the joints and curvature of the TF coils. 

\subsection{Ansys FEA of the SHPD facility design}

For this comparison, one of us (Rana) carried out a 3D FEA analysis of the 2020 design of the Sustained High Power Density (SHPD) facility design\cite{brown_development_2021,menard_fusion_2022,rana_concept_2021}. SHPD is a prospective experiment, designed to research core-edge integration. It is designed to use deuterium plasma, and would not exhibit significant fusion neutron flux. It is a small device, with 1.2 m major radius, aspect ratio 2.4, magnetic field on-axis 5.5 T. The field at the coil surface is 13 T.

A PROCESS model was generated to match the TF coil design of the 2022 version of SHPD, unpublished. The fraction of the vertical separating force on the inner leg (an input, not calculated) was determined by matching the vertical stress at the inboard edge of the midplane of the inner TF coil case. The output files used can be found in the supplemental material.\cite{swanson_supplemental_2022}

The Ansys FEA determined that the point of maximum Tresca stress was the inner surface of the inner TF coil case at the midplane. It was determined to be 705 MPa. The PROCESS model produces a value of 659 MPa, which agrees with the FEA to within 6.5\%.

\subsection{COMSOL FEA of the FESS FNSF facility design}

For this comparison, we examined the 2018 design of the Fusion Energy Systems Studies Fusion Nuclear Science Facility (FESS FNSF)\cite{kessel_overview_2018,zhai_conceptual_2018}. The FESS FNSF is a prospective nuclear science facility, designed to research the nuclear environment of a burning Deuterium-Tritium plasma. It is similar to a pilot plant, but less emphasis is placed on generating net electric power. Its design has 4.8 m major radius, aspect ratio 4.0, magnetic field on-axis 7.5 T, and proposes to produce 518 MW of fusion power. The TF coil design uses Nb$_3$Sn superconductor.

\begin{figure}
\begin{center}
\includegraphics[width=0.9\linewidth]{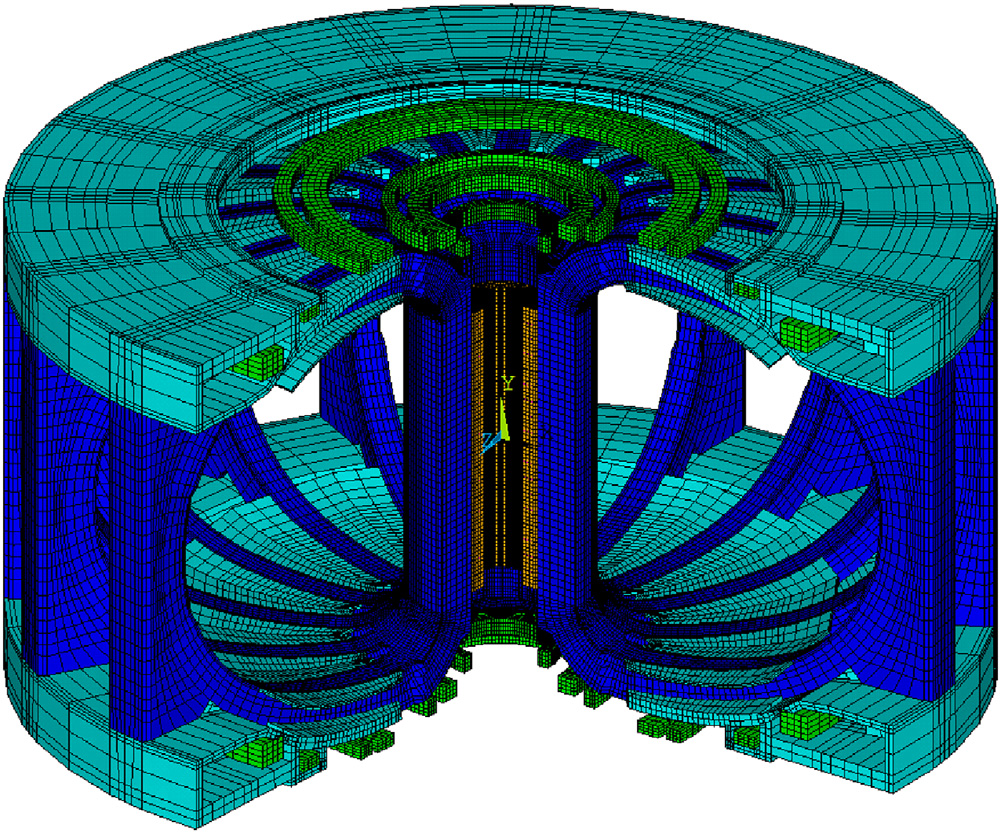}
\caption{The CAD model of FESS FNSF which was analyzed using 3D FEA in Reference \cite{zhai_conceptual_2018}. A PROCESS model of this design was created, and the approximate PROCESS model agrees with the maximum Tresca stress to within 10\%.}
\label{fig:ZhaiFESSFNSF}
\end{center}
\end{figure}

Reference \cite{zhai_conceptual_2018} includes 3D FEA of the FESS FNSF TF coil system. The CAD model that was used is shown in Figure \ref{fig:ZhaiFESSFNSF}. These analyses include features that are not in the SHPD design, allowing more aspects of the approximate model to be validated. The FESS FNSF TF coil system is ``bucked and wedged,"\cite{titus_structural_2003} in which the inner case of the TF coil is radially supported by the outer case of the central solenoid. Furthermore, this paper includes a detailed winding pack analysis which resolves the conduit of the cable-in-conduit conductor (CICC) layout. Because of this, we can validate the smearing rules described in Section \ref{sec:Overview}.

The PROCESS model of FESS FNSF reproduces the core plasma physics parameters to better than 5\%. The output file can be found in the supplemental material.\cite{swanson_supplemental_2022} 

The first point of comparison is at the inner edge of the inboard TF coil case. Reference \cite{zhai_conceptual_2018} reports that FEA produces 600 MPa Tresca stress at this point when bucked and wedged. The approximate model reports 539 MPa Tresca stress at this point, a 10\% underestimate. This 10\% discrepancy is in line with the assumptions that went into formulating the approximate model. The fraction of the vertical separating force on the inner leg (an input in PROCESS, not calculated) was determined using the moment balance method.

The next point of comparison is the maximum Tresca stress of the CICC conduit surrounding the Nb$_3$Sn in the winding pack. Reference \cite{zhai_conceptual_2018} reports this quantity as ``$\sim 1.1$ GPa." The approximate model reports this quantity as 1084 MPa, within the uncertainty of the reported quantity. 

Together, these comparisons validate the approximate model and the ``mixture" or ``smearing" rules that produce bulk effective properties from inhomogeneous winding pack and vice versa. The output of the approximate model may be trusted to 10\% accuracy.

\section{Using PROCESS to optimize a winding pack and coil case}
\label{sec:Optimize}

This section demonstrates the utility of the model by using it to enforce the Tresca yield criterion as a constraint on a many-dimensional optimization of a Tokamak-based fusion pilot plant. This section simply demonstrates that the model can be used in an on-line optimization; it is not designed to argue that the the final design point is advantageous. Because of this, we will not discuss the finer points of the design such as the central solenoid design, power exhaust and divertor design, or quench protection.

Specifically we analyze the Fusion Pilot Plant (FPP) design given in Section 5 of Menard \textit{et al.} (2016)\cite{menard_fusion_2016}. That FPP design is a superconducting, fully non-inductive Spherical Tokamak, designed to produce a nominal amount ($<$100 MW) of net electric power. Details of the radial build of the TF coil are taken from private communication with J. Menard. Details of the HTS winding pack are taken from private communication with Y. Zhai and T. Brown. The conductor section of the winding pack depicted in Figure \ref{fig:wp_decomposition} is assumed to consist of minimally insulated stacked tapes, like that of Tokamak Energy Ltd.,\cite{brittles_hts_2020} so that the axial and transverse Young's modulus may be approximated from that of Hastelloy and known samples.\cite{fujishiro_database_2005}

PROCESS has previously been used to optimize the TF coil system of Tokamak-based power plants. These efforts used a simpler TF coil stress model, which did not include more than 2 layers, self-consistent axial stress and strain, anisotropy, or the contribution to Young's modulus of the winding pack of the conductor itself (significant for REBCO tapes, but not for NbTi or Nb$_3$Sn). Reference \cite{lee_optimal_2015} optimizes a REBCO-based Tokamak. Reference \cite{morris_implications_2015} optimizes an EU-DEMO-like Nb$_3$Sn-based Tokamak. Reference \cite{chislett-mcdonald_could_2019} optimizes NbTi-based Tokamak, finding that the yield stress of the steel conduit and case is not a significant constraint given the low magnetic field strength required by NbTi coils.

A PROCESS model was created for the Menard FPP by following the procedure of Muldrew \textit{et al.} (2020)\cite{muldrew_process_2020}, which models the Fusion Nuclear Science Facility (FNSF) discussed in Sections 2 - 4 of that same J. Menard reference. Selected properties of the PROCESS model of the Menard FPP are given in the ``Original" column of Table \ref{tab:TFOpt}; most parameters match reference \cite{menard_fusion_2016} to better than $5\%$.

\begin{table}
\begin{center}
\begin{tabular}{|p{0.4\linewidth} | p{0.2\linewidth} | p{0.2\linewidth}  |} 
\hline
& Original & Optimized \\ [0.5ex] 
\hline\hline
Major radius & 3.0 m & 3.0 m \\ 
\hline
Aspect ratio & 2.0 & 2.0 \\
\hline
Toroidal field & 4.02 T & 4.61 T \\
\hline
Toroidal current & 12.4 MA & 13.7 MA \\
\hline
Elongation (100\% flux) & 2.49 & 2.49 \\
\hline
Plasma $\beta$ & 7.26\% & 7.00\% \\
\hline
Current drive power (NBI, $E=500$ keV) & 50.0 MW & 50.0 MW \\ 
\hline
Fusion power & 514 MW & 796 MW \\ 
\hline
Net electric power & 52.7 MW & 194 MW \\ 
\hline\hline
Number of TF coils & 12 & 12 \\
\hline
TF coil inner case thickness & 20.0 cm & 37.7 cm \\
\hline
TF coil winding pack thickness & 24.0 cm & 12.0 cm \\
\hline
Total TF coil current & 60.2 MA & 69.1 MA \\
\hline
TF coil current per turn & 10.5 kA & 10.5 kA \\
\hline
TF coil conduit thickness around each turn & 1.5 mm & $\rightarrow 0$ \\
\hline
TF coil turn side length & 12.0 mm & 8.66 mm \\
\hline
TF coil copper fraction per turn & 16.5\% & 16.5\% \\
\hline
Central solenoid outer radius & 23.3 cm & 23.3 cm \\
\hline
Superconductor axial strain & 0.20\% & 0.21\% \\
\hline
TF coil max Tresca stress, case & 667 MPa & 670 MPa \\
\hline
TF coil max Tresca stress, winding pack & 550 MPa & 574 MPa \\
\hline
TF coil operating temperature & 20 K & 20 K \\
\hline
TF coil winding pack current density & 72.7 MA/m$^2$ & 140 MA/m$^2$ \\
\hline
Max $B$ on conductor & 19.2 T & 20.4 T \\
\hline
Critical current fraction $j/j_c$ & 61.8\% & 70.0\% \\
\hline
Temperature margin & 10.6 K & 7.9 K \\
\hline

\end{tabular}
\end{center}
\caption{In the ``Original" column, parameters are given which result from the PROCESS analysis which targets the Menard FPP design point\cite{menard_fusion_2016}. In the ``Optimized" column, parameters are given which result from the PROCESS analysis which starts from that design point, but optimizes for net electric power subject to several constraints detailed in the text. Most notably, net electric power is increased by increasing the toroidal magnetic field, which is enabled by thickening the inboard TF coil case, operating the winding pack closer to its critical current density, and eliminating the steel conduit which is obviated by the Hastelloy cladding of the HTS tapes.}
\label{tab:TFOpt}
\end{table}

As can be seen from ``Original" column of Table \ref{tab:TFOpt}, the original Menard FPP TF coil design is is at $\sim62\%$ of the critical current density, which may be increased while still maintaining significant margin. We used PROCESS to optimize this design for net electric power, subject to many constraints, including the Greenwald density limit, the Troyon beta limit, the LH power threshold, the TF coil case and winding pack conduit Tresca yield criterion (670 MPa), the TF coil conductor critical surface (70\% of $j_c$ for margin), and a limit on the ratio of separatrix power to major radius, $P_{sep}/R$ (20 MW/m). The major radius and aspect ratio were kept constant. The output files used can be found in the supplemental material.\cite{swanson_supplemental_2022}

As can be seen from ``Optimized" column of Table \ref{tab:TFOpt}, PROCESS is able to find a design point with higher net power than the original Menard FPP design point ($52.7$ MW$\rightarrow 194$ MW), which satisfies all the constraints. Primarily this is achieved by increasing the magnetic field, which is enabled by increasing the current density in the winding pack, eliminating the conduit from the winding pack, and increasing the thickness of the inboard steel case. The resultant TF coil case has less conductor, but is operated closer to the critical current density of the HTS. 

The stiffness and strength of the Hastelloy cladding of the HTS tapes are seen to be great boons to the winding pack structure. Because of this strong material, the steel conduit of each turn is unnecessary and is optimized to zero. The current density in the winding pack is increased ($72.7$ MA/m$^2\rightarrow 140$ MA/m$^2$) by operating closer to the critical current ($61.8\%\rightarrow 70\%$) and eliminating the steel conduit (areal fraction $44.4\%\rightarrow 0\%$). Based on Tokamak Energy remarks, 140 MA/m$^2$ winding pack current density at 20 K and 16 T on-coil appears to be within the realm of possibility.\cite{brittles_hts_2020}

Stacked tape TF coil designs face their own specific challenges. They are un-insulated, meaning that it takes a time on the order of their parallel inductive-resistive time $\tau_{L/R,||}$ for the current to be taken up by the superconductor rather than the resistive metal components of the coil. For large fusion magnets, this can be days or even months, necessitating complex charging and terminal design.\cite{titus_startup_2021,brittles_hts_2020} Furthermore, these coils place the HTS tapes themselves under significant transverse stress. The design to which PROCESS optimized, for example, has the HTS tapes under 283 MPa of compressive radial stress, which is high but probably not sufficiently so to cause failure. These specific challenges of stacked tape coil designs are not considered in PROCESS. 


\section{Discussion and conclusion}
\label{sec:Conclusion}

We have presented the benchmarking, validation, and results of an approximate model for the stress and strain distribution within the TF coil of a Tokamak, at the inner midplane. The model is shown to be correct when its assumptions of axisymmetry and generalized plane strain are exactly satisfied. The model is shown to deviate by approximately 10\% when the true 3D geometry is used. The model is suitable for applications requiring fast execution, such as large batch runs of analyses or within an optimizer loop. 

The model is implemented in the PROCESS fusion reactor systems code, which performs constrained optimization on a Tokamak-based power plant design. Using it as a constraint, a Spherical Tokamak pilot plant was optimized for net electric power. The model kept the maximally stressed point of the inner TF coil case below its Tresca yield criterion, and showed that the YBCO's Hastelloy cladding was sufficiently stiff and strong to prevent winding pack yield without steel conduit. 

Some basic dependencies of the Tresca stress on the magnetic field and geometry can be approximately determined from this model. For example, the model places a limit on the magnetic field strength at the surface of the coil, requiring it to be $< 29$ Tesla if a steel TF coil case is used. In practice the limit is significantly below this.

Now that the accuracy of the model has been established ($\sim 10\%$), it may be used as a constraint by PROCESS or other systems codes to optimize other Tokamak-based pilot plants and power plants. In particular, it is well suited to high-field Tokamaks, very large Tokamaks, and Spherical Tokamaks, each of which is receiving attention in the literature.

Future planned updates to the PROCESS TF magnet model include an exploration of how to better include the effect of the lateral side-wall of the coil case, and a model for out-of-plane forces applied by the Poloidal Field (PF) coil system. The 3D FEA of the bluemira open-source multi-fidelity Tokamak systems code will extend these models to a higher level of fidelity.\cite{coleman_blueprint_2019,holloway_new_2021,franza_mira_2021,morris_preparing_2021}

\section{Discussion and conclusion}

We would like to acknowledge the diligence with which Michael Kovari (CCFE) reviewed both the PROCESS implementation of the model and this manuscript. We would like to acknowledge Stuart Muldrew and the PROCESS team at CCFE. 

This work was supported by the U.S. Department of Energy under contract number DE-AC02-09CH11466. The United States Government retains a non-exclusive, paid-up, irrevocable, world-wide license to publish or reproduce the published form of this manuscript, or allow others to do so, for United States Government purposes.

This work has been partly funded by STEP, a UKAEA programme to design and build a prototype fusion energy plant and a path to commercial fusion.

\section{References}

\bibliographystyle{elsarticle-num}
\bibliography{TFStressManuscript}

\end{document}